# Concentration Distribution of Simple Components Reaction Diffusion in one-dimensional linear model

*Zihan Huang[1], Xuewei Yang\*[1]*

(1. Department of Physics, University of Manchester, M13 9PL, UK)

**Abstract**:

The reaction of volatile matter plays an important role in the process of bringing matter from the surface of the planet to the atmosphere. Therefore, by simulating the mixing and chemical reaction process of volatile matter in the atmosphere during volatilization and diffusion from the planet surface, the concentration distribution of different components in the atmosphere can be studied, which is the problem to be solved in this paper. This paper discusses the diffusion and reaction of simple components in one-dimensional scale from the diffusion process of volatile matter and the reaction process in the atmosphere. The diffusion and reaction models of volatile matter were established, and the basis of the model was given.

The volatilization of substances occurs on any planet with suitable surface conditions [1]. During the volatilization of substances on the surface of the planet, when their components contact each other, these substances will undergo chemical reactions [2], which can produce some substances that can stably exist in the atmosphere [3].

These reactions can affect the concentrations of different components in the atmosphere, and then cause changes in the abundance of elements in the atmosphere over a long period of time [4][5].

Therefore, in order to understand this process, it is necessary to establish a model to analyze the chemical reaction of volatile matter when it diffuses in the atmosphere and its impact on the atmosphere.

In this paper, a one-dimensional concentration distribution model of simple components in the atmosphere will be established from the diffusion of volatile matter and the reaction of volatile matters.

# 1 *Methods*

## 1.1 Vapor and Condensation

The vapor and condensation of atmospheric compositions is directly influenced by temperature and pressure. This section is to build a relation between atmospheric composition and surface elements abundance by phase change [6]. In a thermodynamic equilibrium system, there is phase rule as

$$F = C - P + 2 \quad (1)$$

where $F$ is the degrees of freedom, $C$ at here is the number of components and $P$ is the number of phases. The phases composition can be calculated by their phase diagram. A two-component system which contains $A$ and $H_2O$ was considered as an example in this proposal. The amounts of two condensates are fully determined by conservation of mass, the conservation equations for them are:

$$X(A)_{total} = X(A)_{H_2O_{ice}} M_{H_2O_{ice}} + X(A)_{sol'n} M_{sol'n} \quad (2)$$

$$X(H_2O)_{total} = X(H_2O)_{H_2O_{ice}} M_{H_2O_{ice}} + X(H_2O)_{sol'n} M_{sol'n} \quad (3)$$

The amounts of the two phases obey an above simple lever rule. The relations between multi-phases are more complicated but can still be built by modeling base on the lever rule. Then, the mass of liquid and gas of one composition $A$ is determined under a given temperature and pressure, the liquid can be considered as condensed on surface, which remains a part of abundance of elements those in $A$ molecule.

## 1.2 Chemical Reaction and Relation

The reactions take part in atmosphere always follow with photochemical reactions, as they were exposed to much stronger radiation than surface. To build the model from collision, which is the main method the reactions happen at high altitude, the rate of thermal diffusion in gas is need to be defined to calculate the amount of substance at a height [5].

$$\vec{J} = -D\vec{\nabla} n \quad (4)$$

$J$ (mol m$^{-2}$ s$^{-1}$) depends on the speeds of molecule and the gradient in the concentration of the diffusing species. $D$ is the diffusion coefficient. At the turbopause altitude [7], diffusion of molecules is fast as they can be seen as move freely, the above formulation can be written as diffusion flux relate to gradient.

$$J_Z = K_D n \frac{\partial x_i}{\partial z} \quad (5)$$

$K_D$ is a function of height, $D$ depends on temperature and pressure. Since the diffusion process was built, the photochemical model is as following. A three-body collision reaction of components $A$, $B$ and $C$ can be considered as an example

$$A + B + C \rightarrow AB + C$$

There is following formulation to express reaction rate $\eta$, $n_i$ is the number density of species and $k$ is the rate constant of the reaction:

$$\eta = \frac{d(n_{AB})}{dt} = k n_A n_B n_C \quad (6.a)$$

Obviously the rate of reaction is effected by $n_i$, which is related to diffusion $J_Z$. The faster the diffusion is, the lower the density is, the slower the reaction is. If $c_A$ is the concentration of $A$ at a height, the diffusion differential equation contains chemical reaction can be written as:

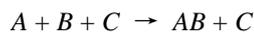

$$c_A \nabla u + \frac{dc_A}{dt} = D_A \nabla^2 c_A + R_A \quad (6.b)$$

where $u$ is the mass average speed of fluid, $R_A = \eta_A$ is the molar average formation rate of $A$.

One common method during atmospheric chemistry researches is to build a chemical reaction net. Now consider an ideal net contains $n$ reactants, $A, B, C \cdots$. There will be $C_n^2$ resultants in the first level for two-body reactions. And the resultants of the second level reactions are:

$$C_{\frac{n(n+1)}{2}}^2 = \frac{n^4 + 2n^3 - n^2 - 2n}{8} \quad (7)$$

Every two-composition reaction has an equilibrium constant $K$ which is only relative to temperature. In the following formulation, $B$ takes part in both two reactions, the relation between two reactions can be built by the partial pressure of $B$, $p(B)$.

$$aA + bB \rightarrow AB \quad , \quad b'B + cC \rightarrow BC$$

$$K_{AB} = \frac{p(AB)}{p^a(A)p^b(B)} \quad , \quad K_{BC} = \frac{p(BC)}{p^{b'}(B)p^c(C)} \quad (8)$$

As the $\eta$ evaluates the degree of reactions and $K$ shows the relations. Then, the amount of a component $A$ at a moment $t$ can be calculated with the reaction rate as it is on the way building an equilibrium. The amount can be expressed by either partial pressure or mole, as original mole plus value changed during $\Delta t$.

$$n_A = n_0 + \sum_k \int_{t-\Delta t}^{t} \eta \, dt \quad (9)$$

$n_A$ is the current mole of $A$ in atmosphere, $n_0$ is the original mole. $k$ is the number of reactions that $A$ takes part in. $\eta$ is the rate of reaction. In each molecule $A$ or $B$ or $C$ contains element $M$ and $N$, which can be denoted as $M_m N$, the abundance of $M$ at this height is $A_{atm}^M$:

$$\frac{\sum n_A \times m}{n_{total}} = A_{atm}^M \quad (10)$$

## 1.3 Volatiles Flux

There are mass exchanges between surface and atmosphere by geological activities like volcanism and weathering both positively [8]. The factors of volcanism were talked in previous work [9]. With considering the light components from a general source, vapor and condensation are mainly considered in this article. As the diffusion coefficient $J$ can be seen as a function of temperature $T$. The temperature is also a function of height $z$, then $J$ is a function of $z$. Consider plate $P_A$ and $P_B$ whose area are both 1m$^2$, the distance between two plates is $\Delta z$ (m). The flux of amount of mass (mol) through the two plates during $\Delta t$ is:

$$\Phi_{P_A}^n = \int_0^{\Delta t} J(z) dt \quad , \quad \Phi_{P_B}^n = \int_0^{\Delta t} J(z + \Delta z) \, dz \quad (11)$$

There is following relation between $\Delta z$ and $\Delta t$. $v_{diff}$ is the velocity of diffusion, which is a function of temperature.

$$\frac{dz}{dt} = v_{diff}(T) \quad (12)$$

Because the temperature is a function of height, $v_{diff}$ can also be seen as a function of height $z$. $p_0$ is the original partial pressure of component $A$ at $z$. Then the average partial pressure of component $A$ in the space formed by plates $P_A$ and $P_B$ is the following formulation, by limit $\Delta z$ as 0, it can be seen as a function of height z finally.

$$p_0 + \frac{\Phi_{P_B}^n - \Phi_{P_A}^n}{\Delta z} = p_M(z) \qquad (13)$$

As $p_M(0)$ is already known, then the partial pressure at different heights can be calculated by the above formulation.

## 2 Results

### 2.1 Conditions Assumption

To simplify the calculation process, 3 light elements, N, H, O were considered to verify the model. Because their components with each other is not as much as C, which is too difficult to the current research situation. They are also not as less as components of Ar, which is not able to test the model. The air composition and temperature with pressure at surface was set as Table.1.

**Table.1** Air composition at surface, 300K, 101MPa

|  | Ratio | Molar mass g/mol |
|---|---|---|
| $N_2$ | 0.8 | 28 |
| $O_2$ | 0.2 | 32 |

The temperature of atmosphere was set as a simple piecewise linear function of height, the pressure was set as a linear function of height and then equal to zero.

T= -6.490×10$^{-3}$ Z + 300K (Z<=10000m)
T= 1.188×10$^{-3}$ (Z-10000) + 216K (10000m<Z<=50000m)
T= -2.400×10$^{-3}$ (Z-50000) + 271K (50000m<Z<=80000m)

P= -8.206Pa/m +101MPa (Z<=20000m)
P=0.1×10$^4$Pa (20000m<Z)

The initial ratio of volatiles on the surface was given in Table.2, which only has reference meaning by giving only two parts of components. The reason to choose them is that their Gibbs formation energy are only two values below 0 in the 7 components of N, H, O in Table.3. In the future this could be changed to observed data. The data of generated molecules during the chemical reaction in atmosphere was listed in Table.3, only these molecules were considered in this article. To simplify calculation, free radicals were not concerned. The diffusion coefficient of each molecule was calculated by (14), where $\sigma_{i-air}$ is the average collision diameter of molecules, $M_i$ is molar mass. Actually, $D_i$ is function of temperature and pressure, then a function of height.

$$D_i = \frac{AT^{\frac{3}{2}}}{P\sigma_{i-air}^2 \Omega}\sqrt{\frac{1}{M_i} + \frac{1}{M_{air}}} \qquad (14)$$

**Table.2** Volatiles at surface, 300K, 101MPa

|  | Ratio | Molar mass g/mol |
|---|---|---|
| $H_2O$ | 0.5 | 18 |
| $NH_3$ | 0.5 | 17 |

**Table.3** Volatiles at surface, 300K, 101MPa

|  | $\Delta_f G_m^{\ominus}$ kJ/mol | $D_i$ cm$^2$/s | d 10$^{-3}$nm | $M_i$ g/mol |
|---|---|---|---|---|
| $H_2O$ | -228.6 | 0.282 | 153 | 18 |
| $NH_3$ | -16.5 | 0.228 | 208 | 10 |
| $H_2$ | 0 | 0.668 | 71 | 2 |
| $O_2$ | 0 | 0.198 | 144 | 32 |
| $N_2$ | 0 | 0.457 | 146 | 14 |
| NO | 87.6 | 0.329 | 119 | 23 |
| $NO_2$ | 51.3 | 0.145 | 285 | 39 |

The collision diameter of air molecules was set as 145,6 (10$^{-3}$ nm). The reaction level was only considered to the second level, which is $C_7^2$ reactions at the first level, $C_{28}^2$ reactions at the second level. In this article, the fluid field was set as a simple vector with module at 10m/s along the vertical direction, the horizontal distribution will not be talked. The total integrate time is the time cost by one mole $H_2O$ generated or diffused to 80000m. The boundary concentrations of mixture of liquid $H_2O$ and $NH_3$ at h=0 were both set as 1mol/m$^3$. The model was run on a one-dimension vertical axis with 20 sampling point.

### 2.2 Concentration Curve

The concentration curve was shown in Fig.1, the NO and $NO_2$ have no notable curve to be compared in a same figure. To $N_2$ and $O_2$, as they are the compositions of air, there are also not notable changes.

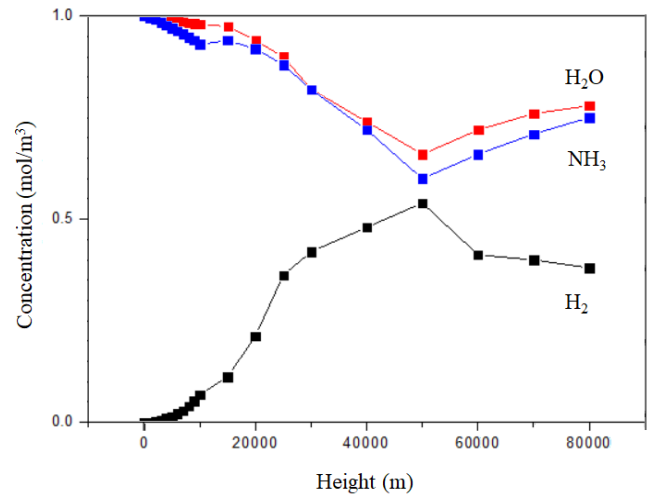

**Fig.1** Concentration curve for $H_2O$, $NH_3$ and $H_2$.

From Fig.1, it can be seen there are strong change trend from 20000 to 40000 meters, which is related to the temperature trend directly. Which is at the height over 20000 meters, the pressure decreases to a very small value, and there is a increasing trend over temperature about 20000 to 40000 meters, which makes it is easier to process reactions with higher Gibbs freedom energy. The most accumulate of substances happens from 40000 to 60000 meters, which is because that the temperature decreases over 50000 meters and reduces the diffusion coefficients of molecules. However, the temperature should not have such influence caused by diffusion on $H_2$ as the concentration decreases 24.78% in 30000 meters in the Fig.1. Although the pressure keeps a very low value, as the temperature decreasing, the direction of reactions tends to the ways which are easier to happen, that makes the concentration $H_2$ start to decrease to form molecules like $H_2O$ and $NH_3$. It can be seen that the concentration of $H_2O$ and $NH_3$ increase over 50000 meters height. The partial pressure curve in Fig.2 gives the similar conclusion, the partial pressure here is the pressure of composition divide the atmospheric pressure.

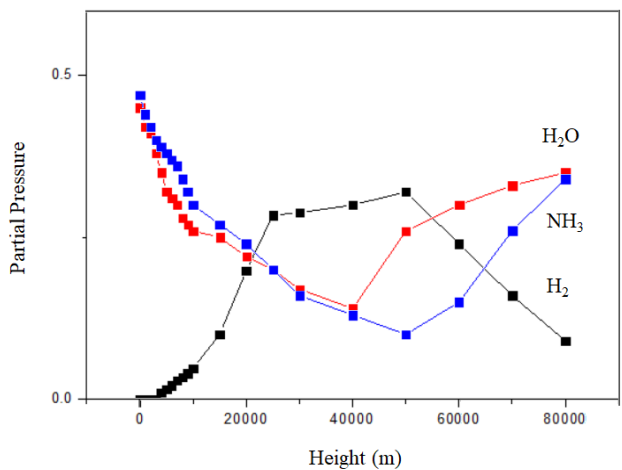

**Fig.2** Partial pressure curve for $H_2O$, $NH_3$ and $H_2$.

## 3 Conclusion

On the basis of previous work, this paper added consideration to chemical reaction in the diffusion process, established the relationship between the concentration of each component through the equilibrium constants of different chemical reactions, and combined with the diffusion mass transfer differential equation to simulate the simple components of three elements in the one-dimensional vertical direction, obtained the potential reaction products and the concentration distribution in the one-dimensional vertical direction.